\documentclass[twocolumn,preprintnumbers,amsmath,amssymb,superscriptaddress, altaffilletter, subeqn, maxbibnames=99]{revtex4-2}

\usepackage{physics}
\usepackage{dcolumn}
\usepackage{bm}					
\usepackage{graphicx}
\usepackage{amsmath}
\usepackage[dvipsnames]{xcolor}
\usepackage{comment}

\begin{document}

\title{Unsupervised learning approach to quantum wavepacket dynamics\\ from coupled temporal-spatial correlations}

\author{Adva Baratz$^*$}

\affiliation{Institute of Artificial Intelligence, Weizmann Institute of Science, Rehovot 7610001, Israel}
\affiliation{Department of Molecular Chemistry and Materials Science, Weizmann Institute of Science, Rehovot 7610001, Israel}

\author{Galit Cohen}
 
\affiliation{Department of Molecular Chemistry and Materials Science, Weizmann Institute of Science, Rehovot 7610001, Israel}

\author{Sivan Refaely-Abramson$^*$}
 
\affiliation{Department of Molecular Chemistry and Materials Science, Weizmann Institute of Science, Rehovot 7610001, Israel}

\begin{abstract}
Understanding complex quantum dynamics in realistic materials requires insight into the underlying correlations dominating the interactions between the participating particles. Due to the wealth of information involved in these processes, applying artificial intelligence methods is compelling. Yet, unsupervised data-driven approaches typically focus on maximal variations of the individual components, rather than considering the correlations between them. Here we present an approach that recognizes correlation patterns to explore convoluted dynamical processes. 
Our scheme is using singular value decomposition (SVD) to extract dynamical features, unveiling the internal temporal-spatial interrelations that generate the dynamical mechanisms. We apply our approach to study light-induced wavepacket propagation in organic crystals, of interest for applications in material-based quantum computing and quantum information science. We show how transformation from the input momentum and time coordinates onto a new correlation-induced coordinate space allows direct recognition of the relaxation and dephasing components dominating the dynamics and demonstrate their dependence on the initial pulse shape. Entanglement of the dynamical features is suggested as a pathway to reproduce the information required for further explainability of these mechanisms. Our method offers a route for elucidating complex dynamical processes using unsupervised AI-based analysis in multi-component systems. 
\end{abstract} 
 
\maketitle


Light-matter interactions in semiconductors generate energetically-excited particles that can relax into long-lived and stable quantum states~\cite{haug2009, cohen2016fundamentals, toyozawa2003optical}. The dynamical mechanisms underlying these processes and their structural dependencies are of great interest for materials-based quantum information and quantum computing~\cite{QuantumComputing, kennes2021moire, rossi2004excitonic, de2021materials}, with the promise of achieving well-defined quantum coherence upon atomistic design~\cite{fogler2014high,katzer2023exciton}.
Organic molecular crystals are important examples of such structurally designable and optically-active semiconductors~\cite{Cudazo2012, Sharifzadeh2013, Kronik2016}, in which low-energy, strongly-bound, and long-lived excited states are populated following a photoexcitation~\cite{Rao2017, refaely2017origins, Sharifzadeh2018}. 
The population of these states is determined by time-resolved scattering dynamics between the energetically-excited particles and lattice vibrations~\cite{Ginsberg2020, Wilson2013, Schnedermann2019, Seiler2021}. These dynamical mechanisms involve simultaneous multi-particle interactions, leading to coupled temporal and spatial correlations~\cite{cohen2024phonon}. Understanding these processes and their structural origins is necessary to produce design principles for effective light-induced quantum coherence. Yet, due to the involved complexity, achieving this goal remains a significant challenge.

Resolving such convoluted dynamical mechanisms within the framework of artificial intelligence (AI) is highly intriguing. In particular, unsupervised data-driven methods are promising for arranging the wealth of information into clusters sharing similar properties to identify the patterns dominating the analyzed processes~\cite{Shan2015, Aggrawal, Scalia2022}.  
This is often achieved through Principal Component Analysis (PCA)~\cite{49b3e4da-9b30-3652-a630-696063b595d6, JOSSE20121869}, a linear dimensionality reduction technique that identifies the variance across each one of the main participating components to generate a set of features, or principal components, ranked in descending order based on their variance. 
While this procedure offers a convenient display of the main dominating features, its underlying assumption is based on the ability to predict the outcome merely through feature variance. However, dynamical processes in materials are determined by internal spatial and temporal correlations resulting from coupling between the various system variables. These correlations are generally not captured in the variance alone, but rather associated with phase relations representing interaction probabilities between the various components. Another data-driven method that does accounts for spatial-temporal correlations is the Dynamical Mode Decomposition (DMD)~\cite{Rowley2009SpectralAO, Schmid2008}. Yet, this method is limited in unfolding coherence patterns in systems exhibiting non-stationary and strong transient behaviours~\cite{Tu2014, Proctor2014DynamicMD}.

\begin{figure*}
\includegraphics[width=1.0\linewidth]{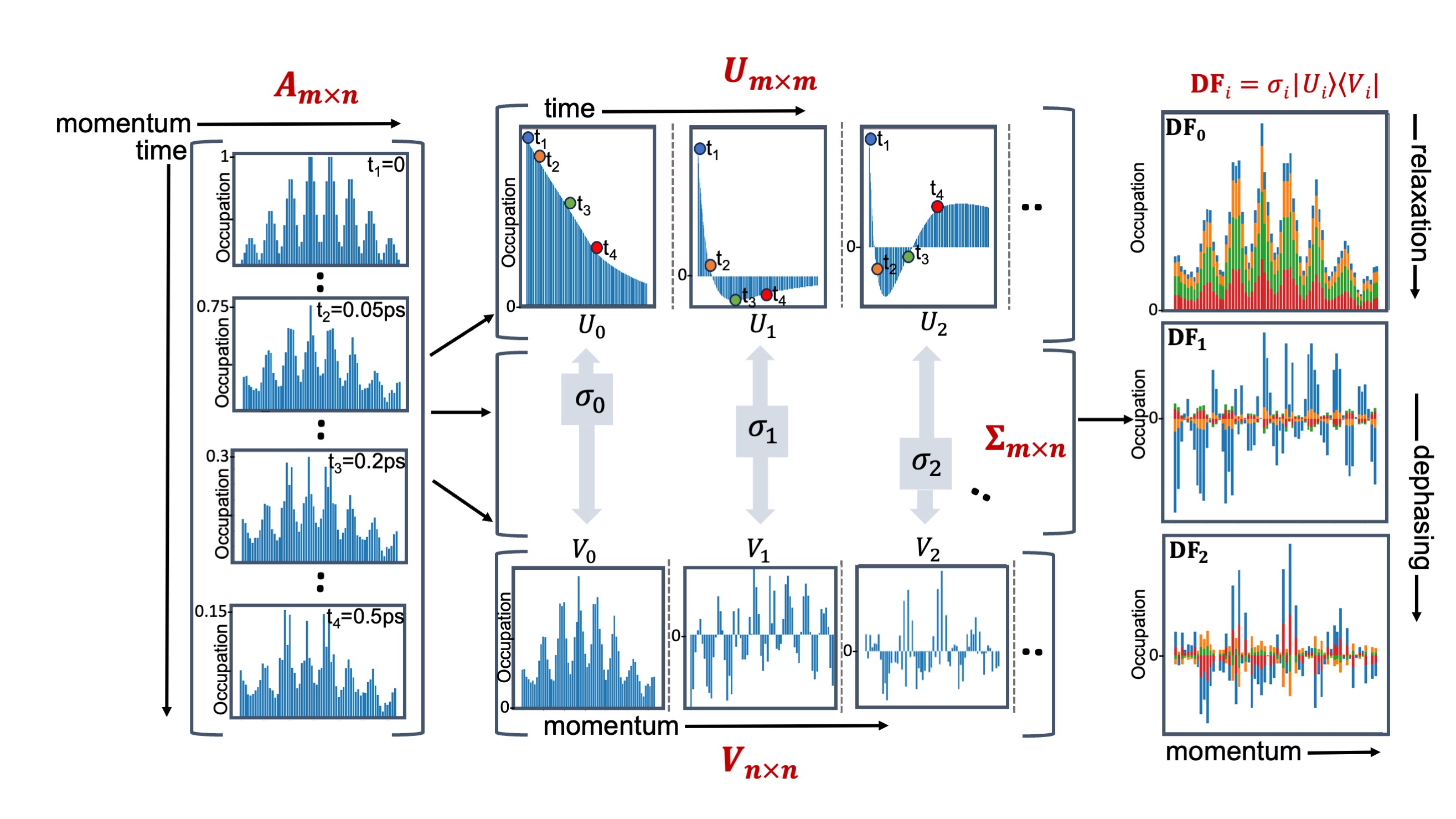}    
    \caption{Left: input matrix $A$ of dimensions $m\times n$ for the occupation dynamics of an initial wavepacket in momentum space, evolving in time ($m$=682 time steps, $n$=64 momentum states). Four representative times in the evolution are presented, between $0$ and $1$~picoseconds (ps). Middle: SVD matrices, $U_{m\times m}$ and $V_{n\times n}$, composed of the singular vectors of the time and momentum coordinates, respectively, and coupled through the diagonal matrix $\Sigma_{m\times n}$ with singular values $\sigma_i$. The four representative times are shown as colored dots upon the evolution of $U$. Right: dynamical features $\mathrm{DF}_i=\sigma_i\ket{U_i}\bra{V_i}$ resulting from the SVD procedure along momentum space with colors associated with the respected times $t_1$-$t_4$. The first dynamical feature, $\mathrm{DF}_0$, contains a gradual decay of the wavepacket amplitude and is associated with relaxation. $\mathrm{DF}_1$ and $\mathrm{DF}_2$ include the first and second component of the phase evolution, associated with dephasing.}\label{fig:svd}
\end{figure*}

A promising general pathway to capture coupled temporal-spatial correlations in the input data is via Singular Value Decomposition (SVD). While this approach is widely used for unsupervised size reduction~\cite{Lay2002-yo, EWERBRING198937}, the numerical possibilities it offers enable analysis that go far beyond mere size reduction, as it transforms the data into a new space that highlights correlations between the participating variables. In this Letter, we introduce an SVD-based approach that exploits the coupling between the left and right singular vectors to uncover the intertwined temporal-spatial correlations dominating complex excited-state dynamics in molecular crystals. We use SVD to transform the input data, initially represented as an evolving wavepacket in the measurement coordinate space, into new coordinates determined by time-resolved and space-resolved interactions. We introduce the resulting correlation-induced dynamical features (DF), which are the operator basis enabling the explainability of the underlying processes determining the involved interaction mechanisms. 
We show that the singular values associated with high-order DFs can be used to quantify the dephasing contribution to the total relaxation as a function of the initial pulse shape. We further introduce the concept of virtual dynamical features (vDF) that extend beyond the classical SVD reconstruction and enable analysis combinations in the space spanned by the singular vectors. Our approach underscores the strength of SVD analysis as a robust explainable data-driven method to study complex particle dynamics in quantum systems.

We demonstrate our method within a propagating wavepacket scheme, a common approach to theoretically describe the occupation distribution of energetically excited states upon laser excitation~\cite{kosloff1994propagation, nitzan2006chemical}, which we recently extended to solid-state systems for studying excited-state dynamics in the pentacene organic molecular crystal from first principles~\cite{cohen2024phonon, Qiu2021}. Changes in the wavepacket occupation distribution along momentum space disclose the propagation dynamics, a result of simultaneous interactions between the involved particles. Figure~\ref{fig:svd} presents the SVD data scheme. Left panel shows the input matrix $A$, with dimensions $m$=682 time steps $\times$ $n$=64 momentum states, containing the computed ab initio data of the wavepacket excitation and propagation (adapted from Ref.~\cite{cohen2024phonon}, see SI for further details). The column and row spaces of $A$ span the evolution of the occupation in time and its distribution in momentum space, respectively. The initial excitation at $t=0$ sets a particle population distribution along $n$ grid points in momentum space, forming the wavepacket with its shape determined by the laser pulse. The evolution of this population in time is demonstrated by the changes in the occupation distribution of $A$ at four selected time steps, from $0$ to $0.5$ picoseconds (ps). The underlying dynamics are determined by two coupled mechanisms: the population relaxation in time and the redistribution of the occupation in momentum space. The former is a result of energy dissipation, expressed in the decay of the wavepacket amplitude. The latter results from dephasing processes, leading to modifications in the wavepacket envelope shape. In the input coordinate representation, these correlations in momentum space are coupled with time and thus are inseparable, and a quantitative understanding of the involved mechanisms is out of reach.

To separate the contributions of these dynamical processes we employ the SVD procedure discussed above, as demonstrated in Figure~\ref{fig:svd} (middle). We decompose the input data onto new coordinate spaces that separate the coupled dynamics through the formation of the orthogonal matrices $U$ and $V$ and the diagonal matrix $\Sigma$ (see SI for further details). The physical interpretation of the new coordinates is dictated by the structure of the input matrix $A$, with rows representing momentum and columns representing time. The columns of $U$ thus set an orthogonal basis for the column space of $A$ comprising independent components of the time dynamics. At the same time, the columns of $V$, forming an orthogonal basis for the row space of $A$, reflect distinct patterns of population distribution in momentum space. Each pattern establishes a specific phase relation among the participating states, where positive and negative values indicate in-phase and out-of-phase relations, respectively. The diagonal elements of $\Sigma$ couple the corresponding left and right pairs of orthogonal vectors from $U$ and $V$ and allow for the estimation of their contribution to the overall dynamics. 

The input matrix $A$ can be fully reconstructed from the SVD decomposition, via:
\begin{equation}\label{Eq:svd_reconstruction}
    A=U \Sigma V^T=\sum_{i=1} ^{\min \{m,n\}} \sigma_i \ket{U_i} \bra{V_i}=\sum_{i} \mathrm{DF}_i
\end{equation}
where we refer to the combined elements $\sigma_i \ket{U_i} \bra{V_i}$ of components $i$ as \textit{Dynamical Features} (DF). Each DF$_i$ characterizes a unique occupation distributions $\ket{V_i}$ in momentum space, forming a phase relation pattern. These distributions undergo a collective evolution, characterized by changes in amplitude and sign, as emphasized by their corresponding dynamical trajectory $\ket{U_i}$. The orthogonality of the columns of $U$ and $V$ leads to the uncoupling of the convoluted dynamics, ensuring that each DF$_i$ evolves independently in space and time.  

Figure~\ref{fig:svd} (right) shows the evolution of the three most significant DFs across time points $t_1$ to $t_4$,  with colors representing the occupation distribution at these times. $\ket{V_0}$ displays the occupation distribution in momentum space, while $\ket{U_0}$ shows the decay of its amplitude over time. Combining them by their singular value $\sigma_0$ gives the overall dynamics of the zero-order DF feature, DF$_0=\sigma_0\ket{U_0}\bra{V_0}$. The evolution of DF$_0$ manifests the correlated \textit{relaxation} of the occupation, which decays in amplitude while preserving its envelope shape.  
The dynamical feature DF$_1=\sigma_1\ket{U_1}\bra{V_1}$, however, is associated with varying phase orientation between the original coordinates as captured in $\ket{V_1}$. Here the coordinates in momentum space are clustered into groups that alternate between positive and negative signs. The dynamics of $\ket{U_1}$ further involve a collective change in sign as it propagates. This feature captures deviations from the shape of the initial occupation distribution, arising from relative changes between populations at different momenta that constitute the wavepacket, thus leading to its \textit{dephasing}. The third feature, DF$_2=\sigma_2\ket{U_2}\bra{V_2}$, is attributed to a higher-order dephasing process. Its dynamics is represented by trajectory with two nodes appearing in its temporal decay. By separating the convoluted wavepacket propagation into these dynamical features, the SVD transformation enables to identify the correlated variations in the occupation distribution, thereby providing access to an intuitive understanding of the dynamical mechanisms involved as well as to quantify their relative contributions.  

 \begin{figure}
\includegraphics[width=0.9\linewidth]{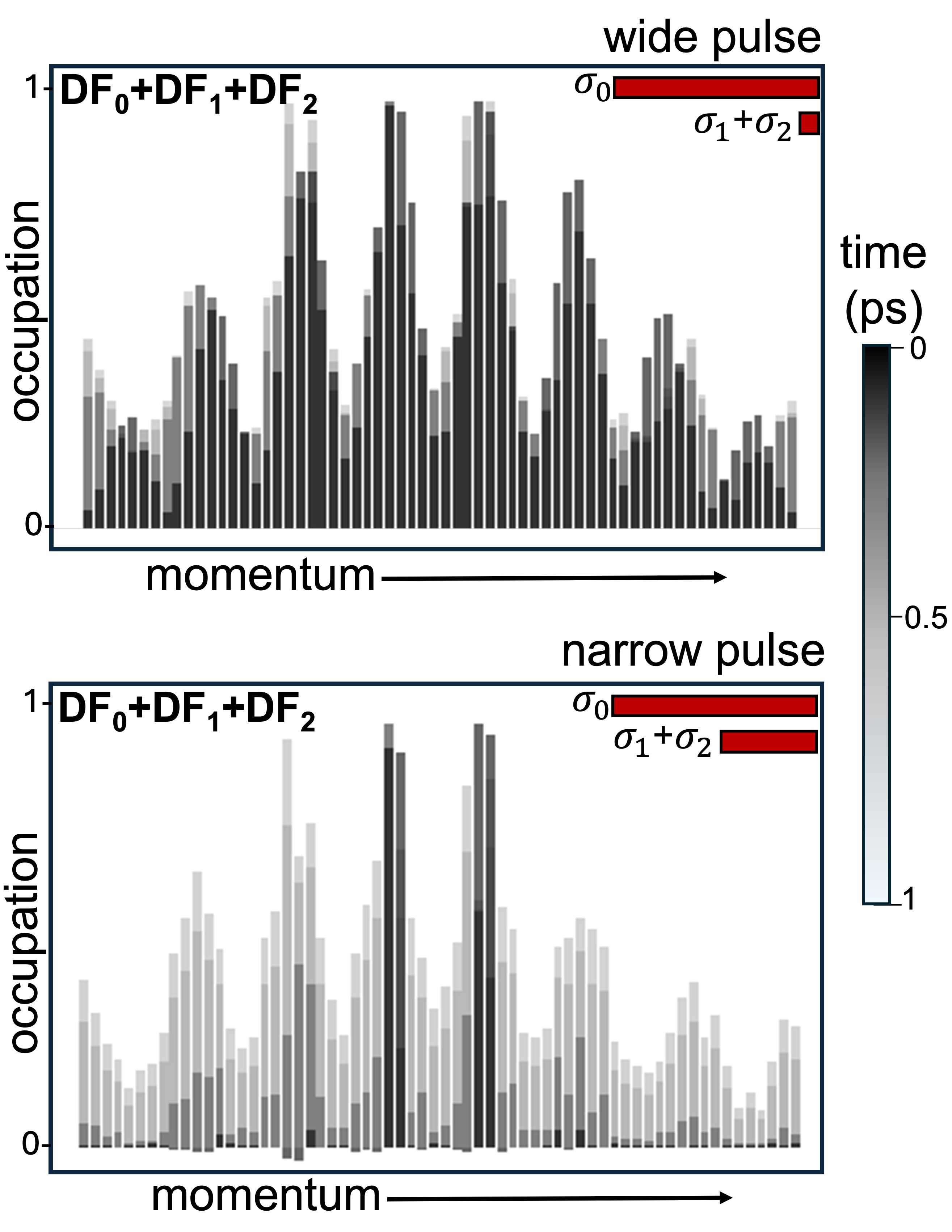}    
    \caption{Time evolution of the normalized wavepacket occupation distribution, from the excitation at $t=0$ to $t=1$~ps, for two different initial conditions initiated by wide (upper panel) and narrow (lower panel) laser pulses. The occupation distribution along momentum space is reconstructed by summing the three largest dynamical features, DF$_0$+DF$_1$+DF$_2$. These different starting points are reflected in the ratio between the singular values $\sigma_i$ of the components associated with dephasing and relaxation, $(\sigma_1+\sigma_2)/\sigma_0$. The wide pulse evolves gradually with the relaxation feature that largely dominates its dynamical mechanism; the narrow pulse generates a strong dephasing component as indicated by the relative increase of $(\sigma_1+\sigma_2)$. Since the crystal structure-- and thus the internal correlations-- are equivalent in both cases, the occupation distribution achieved at a steady state is similar. }\label{fig:sigma}
\end{figure}

The individual contribution of each DF to the overall dynamics is captured by the diagonal elements $\sigma_i$. We demonstrate the physical interpretation of these singular values through an effective reconstruction of the input matrix by tracing the three DFs with most significant contribution, namely $A\approx$ DF$_0$+DF$_1$+DF$_2$. Such reconstruction includes the combined dynamics from the relaxation process captured by DF$_0$ and from the two leading-order dephasing mechanisms captured by DF$_1$ and DF$_2$. Figure~\ref{fig:sigma} shows the evolution of this reconstructed wavepacket from $0$ to $1$ ps, for two different initial pulses: wide (left) and narrow (right). The time evolution follows a redistribution of the occupation among states with different momentum, causing the initial wavepacket to disperse into a less compact shape. The significance of dephasing in the overall dynamics is associated with the ratio $(\sigma_1+\sigma_2)/\sigma_0$, which is $0.1$ for the wide pulse and $0.5$ for the narrow one. It emphasizes the relative weights of the dephasing mechanisms compared to the weight of the relaxation. As this ratio increases, the dephasing contribution diminishes the locality of the initial wavepacket, eventually leading to a normal population distribution in momentum space.
The ratio between the dephasing and the relaxation features indicates the mechanisms leading to a steady state. As the starting point differs substantially from the steady state, the evolved dynamics enforce population redistribution through an enhancement of dephasing processes, as emphasized by their growing contribution to the overall dynamics underscored by a larger $\sigma_i$. 
It is interesting to note that the final wavepacket distribution in momentum space is similar for both pulses, despite the large difference in the initial conditions they imposed. This highlights that while the pulse shape determines the initial excited-state population, the steady state is an internal property of the system and is dictated by its structure, manifested in our case through the correlations along the momentum coordinate. The role of dephasing is thus to drive the system towards this uniform steady state from the, in general arbitrary, starting point.

 \begin{figure}
\includegraphics[width=0.9\linewidth]{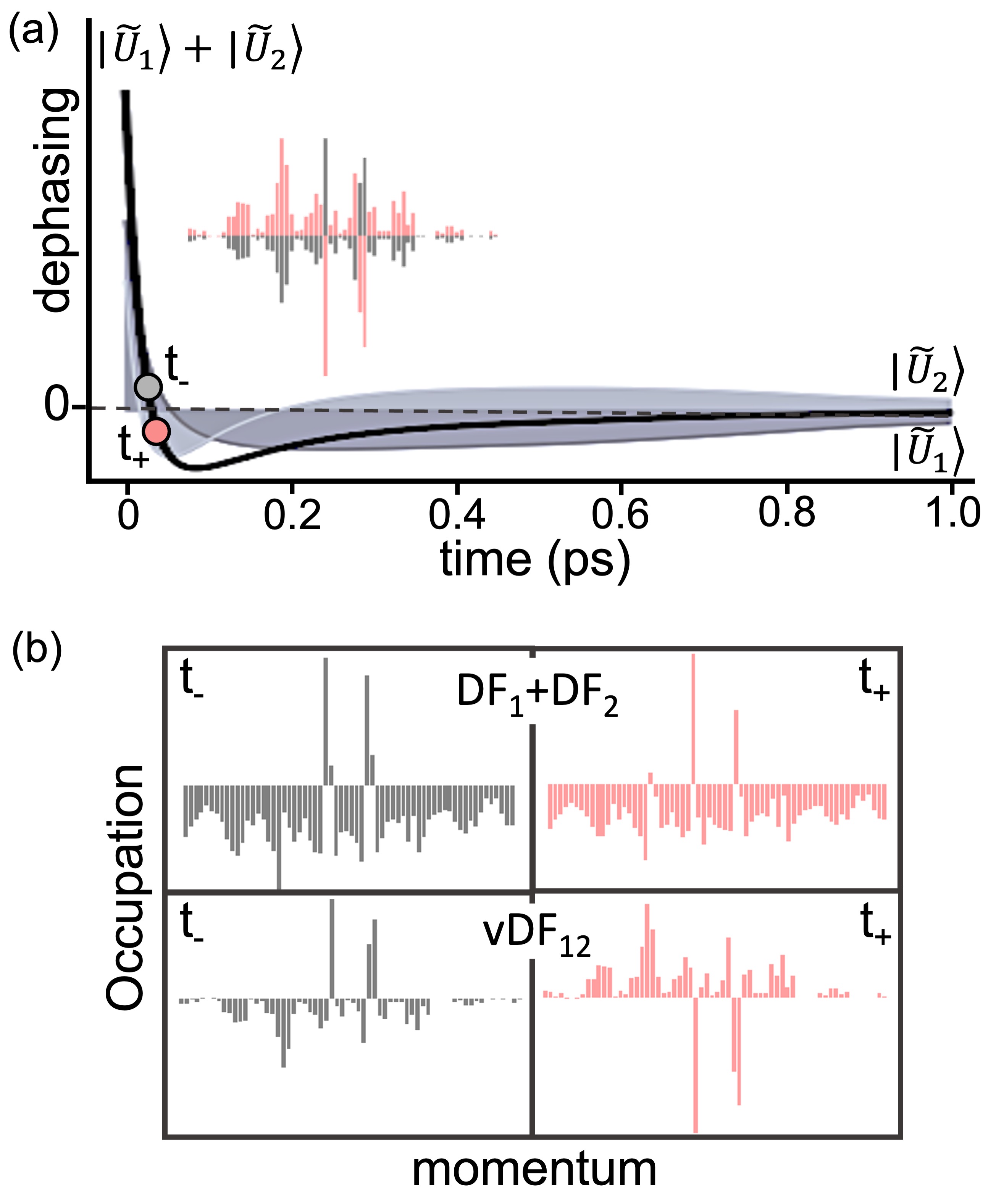}    
    \caption{(a) Time evolution of the dephasing components $\ket{\tilde{U}_1}$ and $\ket{\tilde{U}_2}$ for the two main dephasing features, DF$_1$ and DF$_2$, in the case of a narrow initial pulse. Grey shaded areas show the evolution of the individual components and black line shows the combined evolution of the two, representing the overall dephasing. The combined component asymptotically approaches zero as a steady state is achieved upon mutual cancellation of the individual dephasing components. Two representative times are shown before ($t_-$, grey dot) and after ($t_+$, pink dot) the node of the combined trajectory. (b) Upper panels show the summed distribution occupation of the individual dynamical features, DF$_1$+DF$_2$ at $t_-$ and $t_+$. For the simple summation, a collective sign change before ($t_-$) and after ($t_+$), expected for the mutual component, is absent. Lower panels show the occupation distribution for the entangled virtual feature vDF$_{12}$, for which a collective sign change in the occupation distribution is evident. The same data is also manifested for demonstration as a unified bit in (a), inset.  }\label{fig:df2}
\end{figure}

The case of a narrow-pulse excitation, in which the initial occupation distribution is far from a steady state, results in a substantial dephasing contribution emerging from both DF$_1$ and DF$_2$, with $
\sigma_1=1.4$ and $\sigma_2=0.5$. To study their combined mechanism, it is compelling to analyze the dynamics of the sum DF$_1$+DF$_2$, as presented in Figure~\ref{fig:df2}. The time evolution of the mutual dephasing trajectory is shown in Figure~\ref{fig:df2}(a), with shaded areas depicting the weighted dynamics of the separate components, $\ket{\Tilde{U}_{1,2}} =\sqrt{\sigma_{1,2}}\ket{U_{1,2}}$. Their combined dynamics (black line) are constructed by the sum $\ket{\tilde{U}_1}+\ket{\tilde{U}_2}$. This trajectory decays abruptly at short times, alternates in sign, and gradually stabilizes until asymptotically reaching a steady state with zero dephasing. 
Notably, the combined phase relation of the individual contributions, $\ket{V_1}+\ket{V_2}$, undergoes changes over time and does not maintain a constant pattern throughout the trajectory $\ket{U_1}+\ket{U_2}$. As a result, no collective inversion of the sign in the occupation distribution pattern for DF$_1$+DF$_2$ is evident along the transition between t$_-$ (grey) and t$_+$ (pink), as demonstrated in the upper panels of Fig.~\ref{fig:df2}(b). 

To properly trace the combined phase relation, one can instead account for an entangled state of these two features.
We refer to this state as \textit{virtual dynamical feature} (vDF), defined as:
\begin{equation}\label{eq:vDF}
\begin{split}
    &\mathrm{vDF}_{12}=
    \left(\ket{\Tilde{U}_1}+\ket{\Tilde{U}_2}\right)\left(\bra{\Tilde{V}_1}+\bra{\Tilde{V}_2}\right) =\\
    &(\mathrm{DF}_1+\mathrm{DF}_2) + \ket{\Tilde{U}_1}\bra{\Tilde{V}_2} + \ket{\Tilde{U}_2}\bra{\Tilde{V}_1}
\end{split}  
\end{equation}
for the scaled vectors $\ket{\Tilde{U}_i}=\sqrt{\sigma_i}\ket{U_i}$ and $\ket{\Tilde{V}_i}=\sqrt{\sigma_i}\ket{V_i}$. These features are constructed from a linear combination of the $\ket{{V_i}}$ vectors, effectively forming a new state in the space spanned by the columns of $V$, and the corresponding linear combination of $\ket{{U_i}}$ vectors, effectively forming the associated dynamical trajectory. In practice, such construction first takes into account the mutual phase relation between the momentum occupation distribution before coupling it with the combined time trajectory.  
This virtual feature indeed encapsulates the correlation information, hidden in the initial coordinate space and embedded in the correlation-induced phase relations. As shown in the lower panels of Fig~\ref{fig:df2}(b), the entangled state vDF$_{12}$ captures the collective phase change at the node of the mutual trajectory, preserving the pattern of the occupation distribution along it (further emphasized by overlaying the two time steps in the inset of Fig~\ref{fig:df2}(a)), with the occupation distributions mirroring one another.

To conclude, we introduced an unsupervised learning approach, based on SVD, that captures internal spatial-temporal correlation patterns in complex dynamical processes. By studying the case of quantum wavepacket propagation in organic crystals, we demonstrated the transformation of particle population variations, correlated in space and time, onto collective correlation coordinates that reflect the underlying processes. We exploit the coupling between each pair of left and right singular vectors to reconstruct dynamical features that allow direct recognition of the relaxation and dephasing mechanisms dominating the wavepacket propagation. We further demonstrated how the contribution of these features varies with changes in the initial conditions determined by the shape of the excited pulse, allowing direct connection between pulse shape and the underlying dynamics.
Finally, we presented a construction of virtual features that give access to the phase relations participating in the dephasing process. 
Our method introduces a fully data-driven analysis of complex dynamics without reducing information, using the correlations presented in the input data to identify the natural coordinate space, where the convoluted dynamics is separated into orthogonal mechanisms, allowing for the explainability of the underlying processes.\\

\noindent{Acknowledgments}: we thank Ronnie Kosloff, Tamar Stein and David Zeevi for valuable discussions. A.B. acknowledges support by the Weizmann Artificial Intelligence Institute and Hub. G.C. acknowledges an Institute for Environmental Sustainability (IES) Fellowship. S.R.A. acknowledges a European Research Council (ERC) Grant No.101041159 and an Israel Science Foundation Grant No. 1208/19.

\bibliographystyle{ieeetr}

\end{document}